\documentclass[intlimits,twoside,a4paper]{article}

\usepackage[cp1251]{inputenc}
\usepackage{xcolor}
\usepackage[eqsecnum]{cmpj3}

\usepackage{bm}

\issue{2020}{23}{3}{34601}
\doinumber{10.5488/CMP.23.34601}

\articletype{Rapid communication}

\articletype{Rapid communication}

\title[Water-methanol mixtures]
{On the apparent molar volume of methanol in water-methanol mixtures.
Composition and temperature effects from molecular dynamics study} 

\author[M. Cruz Sanchez, M. Aguilar, O. Pizio]{M. Cruz Sanchez, M. Aguilar,
O. Pizio\thanks{Corresponding author:
\href{mailto:oapizio@gmail.com}{oapizio@gmail.com}}}

\address{Instituto de Quimica, Universidad Nacional Autonoma de Mexico, Circuito Exterior, 04510, \\Cd. de Mexico, Mexico}

\date{Received August 4, 2020, in final form August 21, 2020}
\begin{document}
\maketitle

\begin{abstract}
We explore the effects of composition and temperature on the apparent
molar volumes of species of water-methanol mixtures.
Isothermal-isobaric  molecular dynamics simulations 
are used with this purpose. Several combinations of models for water
and for methanol are explored.  Validity of predictions concerned with a puzzling 
minimum of apparent molar volume of methanol in water-rich solutions is tested
against experimental results. 



\end{abstract}


Experimental studies of aqueous solutions of alcohols and other co-solvents have long
history and continue nowadays. Early research is frequently attributed to
Mendeleev \cite{dima}. Concerning particular systems, the water-methanol liquid mixtures
seemingly are the most studied in various aspects. 
One of the principal properties 
measured and discussed in several works is the behaviour of density on composition,
temperature and pressure. Namely, the density of water-methanol mixtures in the entire
composition interval at normal pressure and various temperatures (e.g. 278.15~K, 288.15~K, 298.15~K,
313.15~K and 323.15~K)  has been reported in \cite{mikhail,mcglashan,sakurai,benson,easteal}.
Equivalently, the mixing properties of two components, methanol and water, can be expressed
in terms of the excess mixing volume, $ \Delta V_{\text{mix}}=V_{\text{mix}}-(1-X_\text{m})V_\text{w}-X_\text{m}V_\text{m}$,
where $V_{\text{mix}}$, $V_\text{w}$ and $V_\text{m}$ refer to the molar volumes of liquid solution and its
components, $X_\text{m}$ denotes the molar fraction of methanol. 
This definition can be rewritten straightforwardly in terms of the respective
densities, see e.g. equation~(2) of \cite{torres}.  It is observed that the
excess mixing volume is negative in the entire composition
interval and exhibits maximum deviation from ideality 
around equimolar composition, at $X_\text{m} \approx$ 0.5.

Moreover, the experimental data can be elaborated in terms of apparent molar volume
for each species~\cite{torres}, $V_{\phi}^{(\text{w})}= V_\text{w} - \Delta V_{\text{mix}}/(1-X_\text{m})$ 
and  $V_{\phi}^{(m)}= V_m - \Delta V_{\text{mix}}/X_\text{m}$.
While, $V_{\phi}^{(\text{w})}$ takes linearly descending values with increasing methanol fraction,
the behaviour of $V_{\phi}^{(\text{m})}$ is peculiar. Namely, if $X_\text{m}$ increases
starting from very small values (i.e. from the limit of water-rich solution), 
 $V_{\phi}^{(\text{m})}$ falls down and  passes through a minimum value, next it grows up
with further growth of $X_{\text{m}}$. The minimum value of~$V_{\phi}^{(\text{m})}$ occurs at $X_\text{m} \approx$ 0.12
for the mixture at room temperature, 298.15~K.   
Similar observations in terms of contraction of the volume of water-methanol mixtures
on composition have been discussed recently in~\cite{malomuzh1,malomuzh2}.
 
Peculiar points of the behaviour of properties of water-methanol mixtures along 
composition axes and at different temperatures are also drawn as a logical conclusion 
from the analyses of light scattering experimental data, see e.g.~\cite{parfitt,micali}. 
A qualitatively similar picture involving the apparent molar volume behaviour with trends of
light scattering experimental data has been discussed for water-DMSO solutions in~\cite{rodnikova}.
It is quite clear that the peculiarities of mixing of species in which one component 
consists of molecules with hydrophilic and hydrophobic parts (besides, 
both species are able to form hydrogen bonds) result from changes of intermolecular structure
upon changing composition and temperature.

After these preliminaries, we would like to establish the principal objective of the present 
brief report. Specifically, our focus is to make clear if computer simulations of models
for water and methanol are able to describe fine details of some properties observable 
from experimental data of this type of mixtures. Previously, there have been very many
computer simulation study of water-methanol mixtures. For the sake of convenience of the
reader we refer to the list of references in our recent works on these mixtures~\cite{cruz1,galicia1}.

Common statement of the problem includes the choice of the model for water and for methanol
complemented by using certain combination rule for the crossed interactions.
Usually, the Lorentz-Berthelot (LB) combination rules (CR2 in the Gromacs nomenclature) or the
CR3 geometric rules are applied. 

Alternative strategy for a single-component fluid or a mixture does not rely on the 
combination rules. In this framework, it is
assumed that the parameters for the cross-interactions of the given functional form 
follow from the fitting procedure to yield the  target properties known for example, 
from the experiment.  The choice of target properties is not simply the question of taste, 
actually it depends on the focus of specific research and determines range
of applicability of the designed model. Different realizations of this type
of procedure have been presented in \cite{nezbeda1,nezbeda2,bopp,vega-2016,lomba,salas,spohr,zeron}.
\begin{figure}[!b]
\begin{center}
\includegraphics[width=5.5cm,clip]{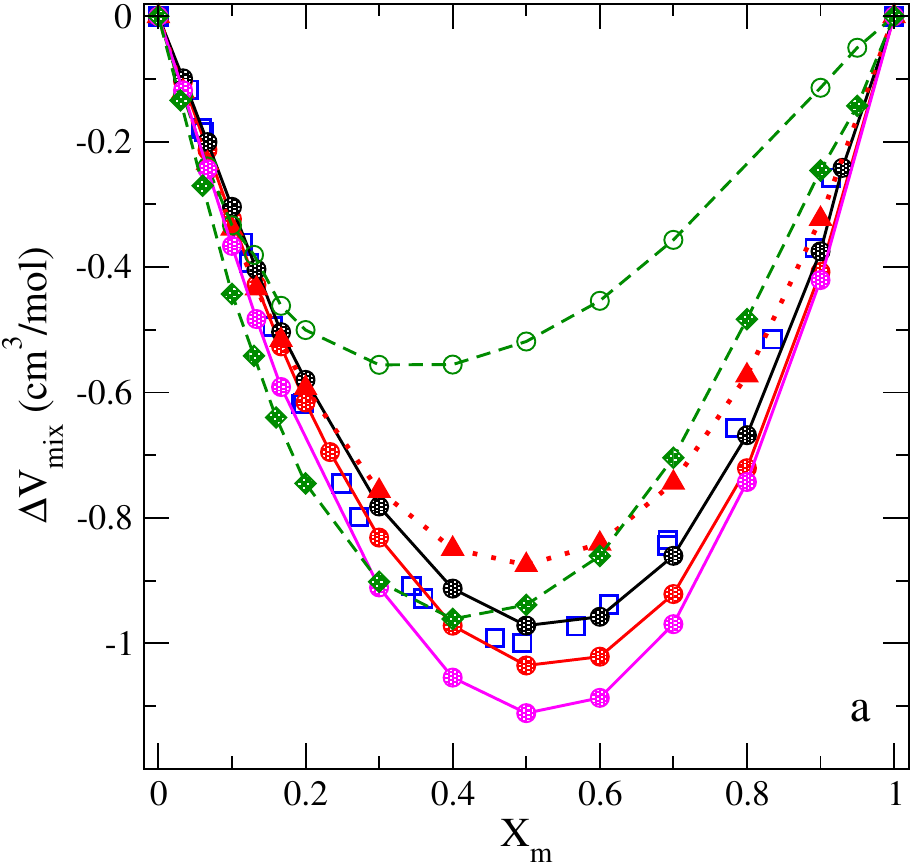}
\includegraphics[width=5.5cm,clip]{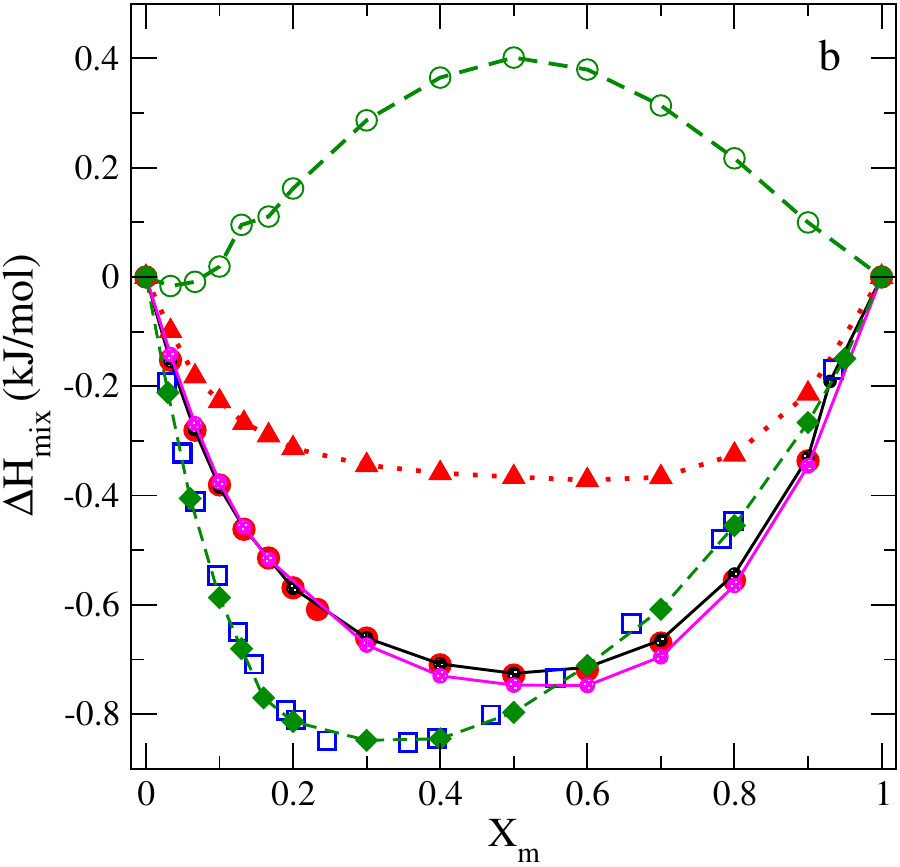}
\end{center}
\caption{(Colour online)
Panel a: Excess mixing volume of water-methanol mixtures on methanol molar fraction
for different combinations of models together with the experimental data (blue squares).
The nomenclature of symbols is the following:
TIP4P-2005-L1 (black circles), TIP4P-2005-TraPPE (red circles), TIP4P-2005-OPLS (magenta circles),
SPC-E-TraPPE (red triangles), TIP4P-2005-OPLS-2016 (green hollow circles) and 
TIP4P-2005-OPLS-2016-F (fitting with parameters from \cite{lomba}, green diamonds).
Panel b:  Excess mixing enthalpy of water-methanol mixtures on methanol molar fraction
for different combinations of models together with the experimental data (blue squares).
The nomenclature of lines and symbols as in panel a.
}
\label{fig1}
\end{figure}

The protocol of simulation of the systems with three thousand of particles 
using Gromacs software (version 5.1) under the 
isobaric-isothermal conditions of the present work is the same as in our previous 
study~\cite{cruz1}. We restrict our attention to nonpolarizable models only.
The results of simulations refer to the TIP4P-2005~\cite{vega-tip} and the 
SPC-E \cite{spce} water models. 
It is known that the the TIP4P-2005 version provides reasonable performance of the
microscopic structure of water~\cite{pusztai} and a quite good description of various
properties, in comparison to other water models, see e.g. table 2 of \cite{vega-pccp}.

On the other hand, a set of united type methanol models has 
been explored. Specifically, we consider three models conveniently described in table~I
of \cite{vega-2016}: L1~\cite{leeuwen}, OPLS~\cite{jorgensen} and OPLS-2016~\cite{vega-2016}.
Besides, the TraPPE methanol model~\cite{trappe} is incorporated. For all these cases
standard strategy is applied. If the OPLS-type modelling is involved, then the CR3 rules
are used. Otherwise, the LB rules determining the cross interactions are applied.
On the other hand, it appeared that the parametrized version of water-methanol 
mixture~\cite{lomba} is helpful for the purposes of our study, as explained below.


  A set of results concerning the excess mixing volume of the models of the present
study at normal pressure, 1 bar,  and at room temperature, 298.15~K, is given in figure~\ref{fig1}~a.
Formally, all the models under exploration yield a qualitatively correct result 
describing the presence of a minimum of $\Delta V_{\text{mix}}$, in accordance with the
experimental data. However, the TIP4P-2005 water combined with L1, TraPPE and OPLS
methanol models all provide the minimum at a correct composition value, $X_\text{m} \approx$ 0.5,  
and predict $\Delta V_{\text{mix}}$ value at minimum reasonably well. The best predictions
are in fact provided by the TIP4P-2005/L1 and TIP4P-2005/TraPPE models. 
In contrast, the SPC-E/TraPPE potential set underestimates the magnitude of the $\Delta V_{\text{mix}}$
at minimum. The curve coming from the TIP4P-2005/OPLS-2016 model exhibits very pronounced
deviation from the experimental values in the entire composition interval,
in spite of the fact that the models for each
individual species apparently  are the best. In our opinion,  the reason  of such
discrepancy lies in the apparent  ``conflict'' between the combination rules and the
methodology of the design of the OPLS-2016 model for methanol. Namely, the
parametrization of the OPLS-2016 model has been performed taking the fluid-solid
equilibrium as a target~\cite{vega-2016}. It has been commented that 
better global performance of the methanol modelling is expected from the inclusion 
of the so-called polarization term~\cite{vega-2016}, similar observation has been 
advertised in~\cite{nezbeda3} while considering water-methanol mixture.

Having in mind unsatisfactory predictions of the TIP4P-2005/OPLS-2016 modelling for 
$\Delta V_{\text{mix}}$ within standard procedure, it has been proposed to apply fitting,
according to the alternative procedure with the excess mixing enthalpy and the excess
mixing volume as target properties. In the case of the excess mixing enthalpy, $\Delta H_{\text{mix}}$
(defined as $ \Delta H_{\text{mix}}=H_{\text{mix}}-(1-X_\text{m})H_{\text{w}}-X_{\text{m}}H_{\text{m}}$), 
fitting leads to a curve perfectly coinciding with experimental data in the entire interval
of $X_\text{m}$, figure~\ref{fig1}~b, in contrast to wrong predictions within  standard procedure. 
On the other hand, the excess mixing volume, $\Delta V_{\text{mix}}$, after fitting, see the green line with solid diamonds in figure~\ref{fig1}~a, still deviates from the experimental data
predicting minimum at a lower $X_\text{m}$ than required. 
Rather accurate description of $\Delta V_{\text{mix}}(X_\text{m})$ has been obtained within
different, microscopic structure demanding, fitting procedure in \cite{bopp}. Unfortunately,
the excess mixing enthalpy has not been reported.
Final comment concerning figure~\ref{fig1}~b is that the TIP4P-2005, if combined with either L1, TraPPE
or OPLS within standard procedure, is of superior quality comparing to SPC-E/TraPPE, 
but does not describe well the magnitude of $\Delta H_{\text{mix}}$ at minimum and its position 
along composition axes.  In close similarity to these developments,
Nezbeda and co-workers~\cite{nezbeda1,nezbeda2} have explored $\Delta V_{\text{mix}}(X_\text{m})$ and
$\Delta H_{\text{mix}}(X_\text{m})$ at room temperature applying adjustment of the parameters for 
the cross interactions and obtained certain improvement of the results, in comparison 
to standard modelling with LB combination rules. However, in summary one can conclude that
the energetic and geometric aspects of mixing of species are intrinsically coupled and
it is difficult to describe both of them perfectly well simultaneously within 
nonpolarizable modelling. 
\begin{figure}[!b]
\begin{center}
\includegraphics[width=5.5cm,clip]{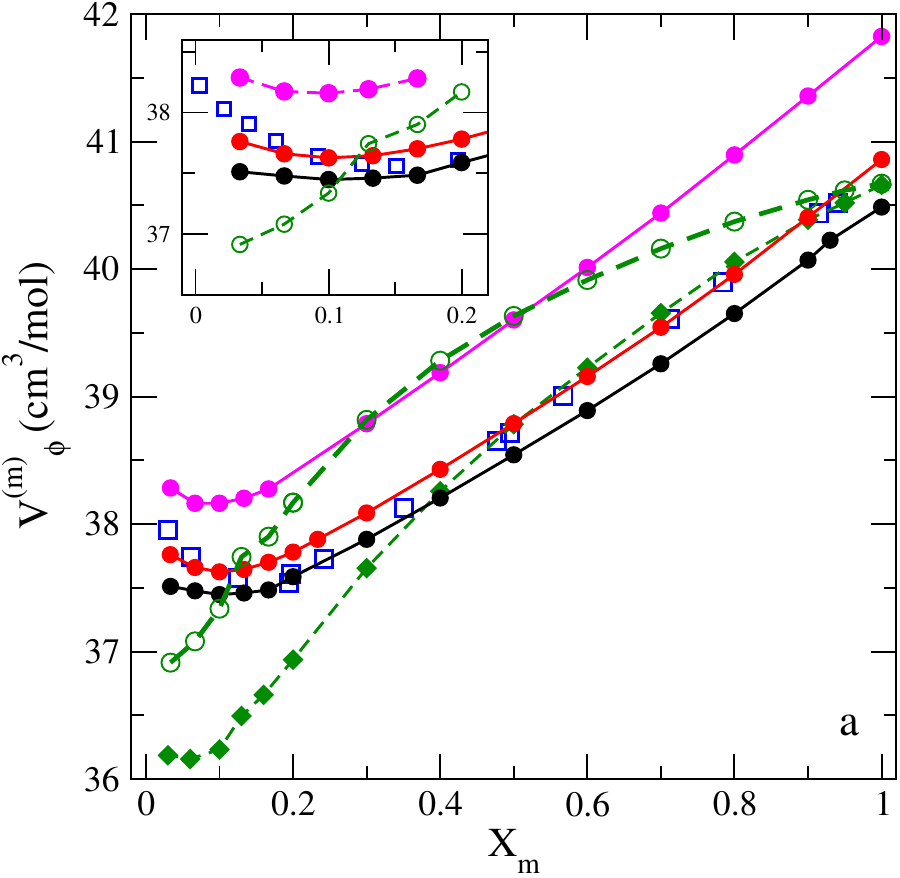}
\includegraphics[width=5.5cm,clip]{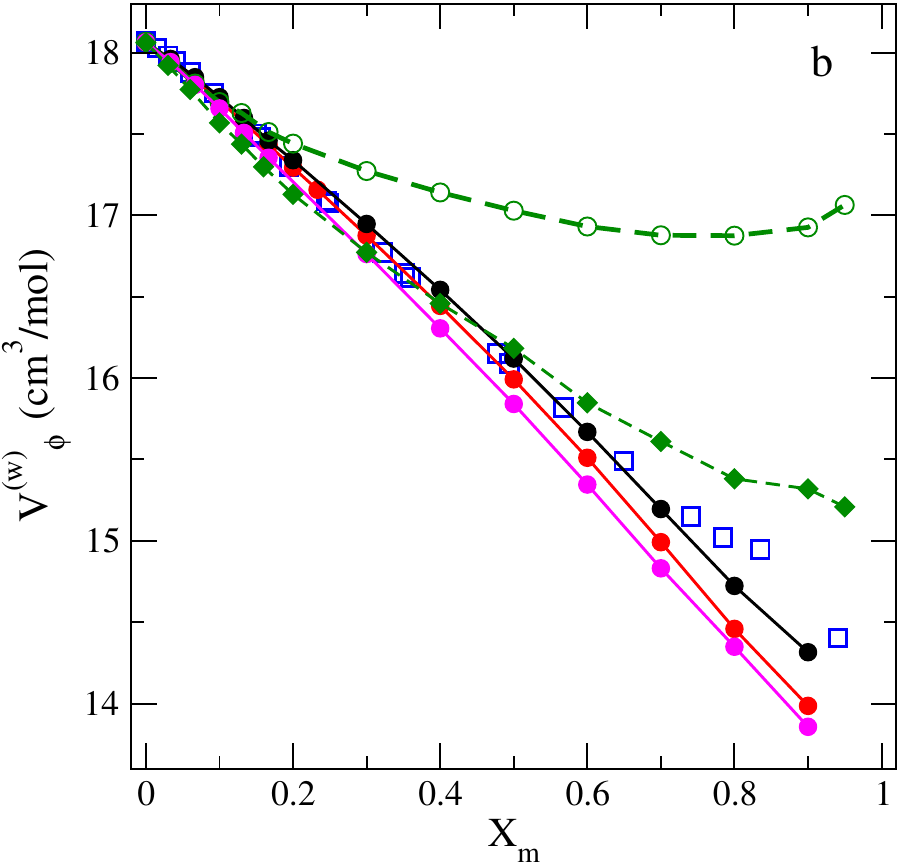}
\end{center}
\caption{(Colour online) 
Apparent molar volumes of methanol and of water (panels a and b, respectively)
for water-methanol mixtures on methanol molar fraction
for different combinations of models together with the experimental data (blue squares).
The nomenclature of lines and symbols as in figure~\ref{fig1}.
}
\label{fig2}
\end{figure}

It is worth examining the available results for $\Delta V_{\text{mix}}$ more rigorously
expecting  to find its possible peculiar behaviour 
on $X_\text{m}$ at a low methanol mole fraction. For this purpose, we have evaluated the 
apparent molar volume for each component, defined at the beginning of this brief report, 
of the water-methanol mixture in the entire composition interval.  
A set of results following from the application of different models is given in 
two panels of figure~\ref{fig2}. Namely, in figure~\ref{fig2}~a, the results for $V_{\phi}^{(\text{m})}$ are shown 
for the models in which TIP4P-2005 water is combined with L1, TraPPE and OPLS methanol. 
The experimental data are given as well. We observe that all three combined models
capture the presence of a minimum of the apparent molar volume of methanol at a low value
of $X_\text{m}$ in agreement with experimental data. Concerning the position of the minimum along
$X_\text{m}$ axes, the simulations predict slightly lower value for $X_\text{m}$ comparing to the 
experiment. The value for $V_{\phi}^{(\text{m})}$ at minimum is predicted by the TIP4P-2005/TraPPE
model best. For this particular model, the entire curve describing $V_{\phi}^{(\text{m})}(X_\text{m})$
is in very good agreement with experimental results. The shape of the curves coming
from TIP4P-2005/L1 and from TIP4P-2005/OPLS is satisfactory as well, but the absolute values
for $V_{\phi}^{(\text{m})}(X_\text{m})$ are less accurate because the results for pure methanol
within these two models slightly deviate from the experiment, in contrast to TIP4P-2005/TraPPE.
Interestingly, the combined model TIP4P-2005/OPLS-2016 does not yield correct behaviour
of $V_{\phi}^{(\text{m})}(X_\text{m})$ at low values for $X_\text{m}$ in the framework of modelling using the
combination rules. If one applies the parametrization according to~\cite{lomba}, then
the minimum at low $X_\text{m}$ is captured, but the absolute values for $V_{\phi}^{(\text{m})}$ are less
accurate, in comparison to other models mentioned above.
According to the authors of recent~\cite{bopp} concerned with the parametrized version
of the original BJH and PHH flexible models for water and methanol,
respectively~\cite{bjh-water,palinkas}, the minimum of  the partial volume of methanol 
at low concentration has been reproduced in that work for the first time.
To be fair, it is worth mentioning that this minimum has been obtained in \cite{nezbeda1},
by using a particular set of parameters for the cross interactions in the TIP4P/OPLS model.
Nevertheless, the $X_\text{m}$ value, and $V_{\phi}^{(\text{m})}$ at minimum in both contributions~\cite{nezbeda1,bopp},
are less accurately predicted, in comparison to the TIP4P-2005/TraPPE modelling of the present report.

Water, in contrast to methanol, does not exhibit peculiar behaviour on the methanol mole fraction
in water-methanol mixtures in terms of $V_{\phi}^{(\text{w})}$. The relevant results are shown in figure~\ref{fig2}~b.
Here, it is important to note that the TIP4P-2005, combined with either L1, TraPPE
or OPLS with combination rules describes the $V_{\phi}^{(\text{w})}$ quite accurately in the entire
composition interval. On the other hand, the TIP4P-2005/OPLS-2016 model is not adequate in this aspect.
Solely parametrization of the cross interactions leads to certain improvement of the results.
The curves for $V_{\phi}^{(\text{w})}(X_\text{m})$, reported in~\cite{nezbeda2,bopp}, are satisfactory with
and without respective parametrizations.
\begin{figure}[!b]
\begin{center}
\includegraphics[width=5.5cm,clip]{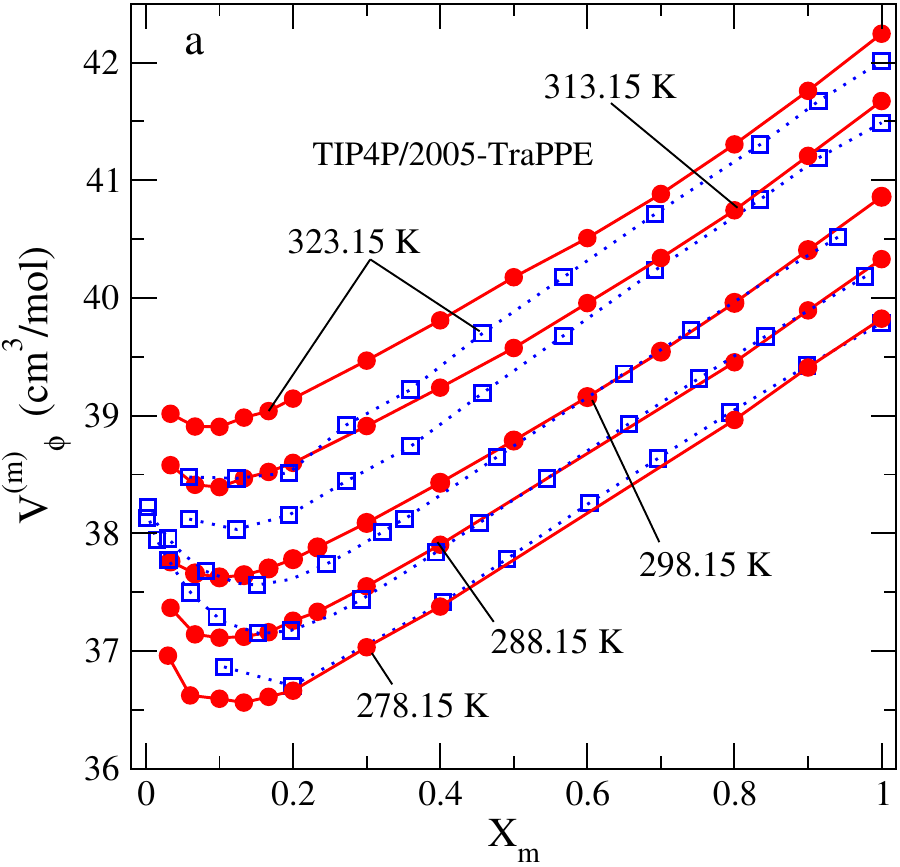}
\includegraphics[width=5.5cm,clip]{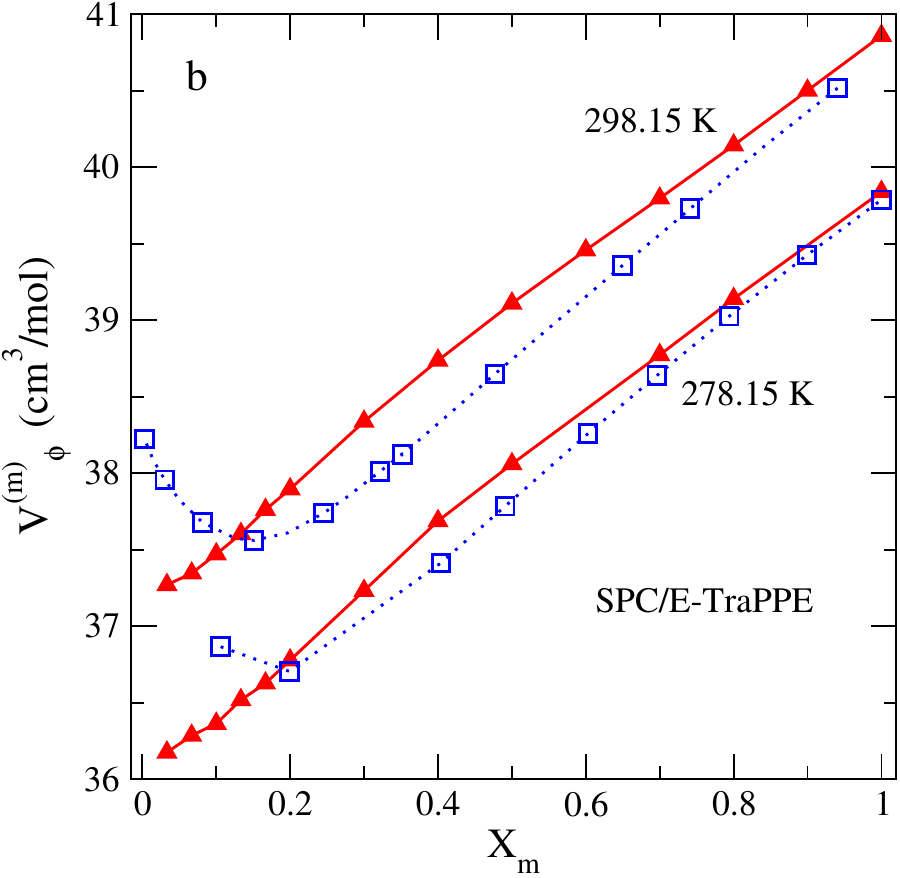}
\end{center}
\caption{(Colour online)
Temperature dependence of the apparent molar volume of methanol
for water-methanol mixtures (panel a: TIP4P-2005-TraPPE model; panel b: 
SPC-E-TraPPE model) on methanol molar fraction
for different temperatures, in comparison to the experimental 
data (blue squares).
The nomenclature of lines is given in the figure~\ref{fig1}.
}
\label{fig3}
\end{figure}

Finally, we would like to comment on the temperature trends for the observed peculiar behaviour.
The experimental data for density of water-methanol mixtures have been reported in the 
temperature interval from 278.15~K to 323.15~K. The methanol molecule contains hydrophobic group,
CH$_3$, besides the OH entity able to participate in the methanol-methanol and methanol-water hydrogen bonding.
Thus, one should expect changes in the balance between hydrophilic and albeit weak hydrophobic effects.
The former would be expected to weaken as temperature increases while the latter formally could possibly drive
hydrophobic aggregation, as it has been discussed, for example, for water-DMSO mixtures~\cite{samios}.
As it follows, from the experimental data for density, the minimum of the apparent molar volume of
methanol, $V_{\phi}^{(\text{m})}$, becomes less pronounced with increasing temperature from 278.15~K to
323.15~K, just a flat fragment is observed at low values of $X_\text{m}$ at the highest temperature studied,
figure~\ref{fig3}~a. The data from simulations of TIP4P-2005/TraPPE model follow experimental trends.
Namely, the magnitude of $V_{\phi}^{(\text{m})}$ increases at each fixed composition, if the temperature increases,
i.e., if the system becomes less non-ideal.
This behaviour is characteristic for the entire composition interval. 
The minimum for the $V_{\phi}^{(\text{m})}$ from simulations is less pronounced at 323.15~K comparing to the
lowest temperature in question, 278.15~K. Moreover, the minimum slightly shifts to a lower value of $X_\text{m}$
upon increasing temperature, similar trend can be deduced from the available experimental data.
On the other hand, accuracy of computer simulation predictions deteriorates, if the temperature increases.
Still, the predictions from the TIP4P-2005/TraPPE model are quite satisfactory. General trends of behaviour
of $V_{\phi}^{(\text{m})}$ with temperature are predicted by the SPC-E/TraPPE water-methanol model, 
figure~\ref{fig3}~b. However, this kind of modelling has serious deficiency --- peculiarity of the $V_{\phi}^{(\text{m})}$
at low $X_\text{m}$ is not captured at all, at two temperatures studied. Possibly, one needs to resort to a
parametrization of crossed interactions similar to \cite{lomba} as we discussed above.
The dependencies of $V_{\phi}^{(\text{w})}$ on $X_\text{m}$ are weakly sensitive to temperature,
all of them behave as declining lines for different models, cf. figure~\ref{fig2}~b.

To conclude, we have performed a wide set of computer simulations for different models of
water-methanol mixtures and analyzed their feasibility to capture peculiarity of mixing of
species in water-rich region. We have resorted to a standard procedure of modelling by
picking up model for each component and apply the combination rules. It has been established 
that the TIP4P-2005 water model combined with either L1 or TraPPE or OPLS models of methanol
reproduce experimentally observed minimum for the apparent molar volume of methanol
at a low methanol mole fraction. The best performance is observed while using the TIP4P-2005/TraPPE 
model. Almost perfect description of this peculiar behaviour is due to a very good description of the
slope of the mixing volume following from correct evolution of the density of the system.
Moreover, analyses of the evolution of model structure, in terms of the pair
distribution functions, coordination numbers and the numbers of hydrogen bonds, 
evidence a progressive mixing of components with increasing $X_\text{m}$  and do not indicate any 
segregation trends similar to the model studied in~\cite{bopp}
(the corresponding figures are not presented due to the format of this communication).
Interestingly, one can attempt to investigate partial properties related to the enthalpy of mixing.
The results of the present study permit to choose adequate models with the aim of 
exploration of properties coming from vibrational spectroscopy experiments and of refractive index,
for example.
\section*{Acknowledgements}
M. Cruz is grateful to CONACyT for support of the Ph.D. studies.

\ukrainianpart

\title[Water-methanol mixtures]
{Видимий молярний об'єм метанолу в сумішах вода-метанол.
Дослідження впливу складу і температури  методом молекулярної динаміки} 

\author{M. Круз Санчес, M.  Агіляр,
O. Пізіо}

\address{Iнститут хiмiї, Нацiональний автономний унiверситет м. Мехiко, Мехiко, Мексика
	}

\makeukrtitle 

\begin{abstract}
Ми досліджуємо вплив складу і температури на видимі молярні об'єми сортів сумішей вода-метанол.
Для цього використано симуляції методом молекулярної динаміки в ізотермічно-ізобаричному ансамблі. 
Досліджено декілька комбінацій моделей води і метанолу. Якість передбачень, пов'язаних із  загадкою мінімуму 
видимого молярного об'єму метанолу у водних  розчинах, перевіряється порівнянням з експериментальними  результатами.

\keywords  суміші вода-метанол,  надлишковий об'єм змішування, видимий молярний об'єм, симуляції молекулярної динаміки


\end{abstract}
\end{document}